\documentclass{ws-procs961x669}            
\begin{document}
\title{Type Ia Supernovae and their Explosive Nucleosynthesis: Constraints on Progenitors}

\author{Shing-Chi Leung$^*$}

\address{TAPIR, Walter Burke Institute for Theoretical Physics, Mailcode 350-17, Caltech, Pasadena, CA 91125, USA\\
$^*$E-mail: scleung@caltech.edu\\
sites.google.com/view/scleung}

\author{Ken'ichi Nomoto}

\address{Kavli Institute for the Physics and 
Mathematics of the Universe (WPI), The University 
of Tokyo Institutes for Advanced Study, The 
University of Tokyo, Kashiwa, Chiba 277-8583, Japan \\
E-mail: nomoto@astron.s.u-tokyo.ac.jp}

\begin{abstract}

  What the progenitors of Type Ia supernovae (SNe Ia) are, whether they
are near-Chandrasekhar mass or sub-Chandrasekhar mass white dwarfs, has
been the matter of debate for decades. Various observational hints are
supporting both models as the main progenitors. In this paper, we 
review the explosion physics and the chemical abundance patterns
of SNe Ia from these two classes of progenitors. We will discuss how
the observational data of SNe Ia, their remnants, the Milky Way Galaxy, 
and galactic clusters can help us to determine the essential features
where numerical models of SNe Ia need to match.
\end{abstract}


\bodymatter

\section{Introduction}
\label{sec:Introduction}

Type Ia supernovae (SNe Ia) are well-understood as the thermonuclear
explosions of carbon-oxygen white dwarfs (CO WDs)\citep{Arnett1969,
  Hillebrandt2000, Nomoto2017SNIa}. They produce the majority of
iron-peak elements in the galaxy, in particular $^{55}$Mn. Their light
curves can be standardized for measuring distance in the cosmological
scale \citep{Riess1998, Perlmutter1999}. Understanding their 
progenitors, the explosion mechanisms and their obseravables are
important for understanding the Universe in the larger scale
\citep{Matteucci2001, Kobayashi2011}. In this review paper, we will
explore possible progenitors of SNe Ia, whether they are the
explosions of near-Chandrasekhar mass (Ch-mass) WDs or
sub-Chandrasekhar mass (subCh-mass) WDs. In Table
\ref{table:compare_WDs} we tabulate the important features to contrast
between the Ch-mass and subCh-mass WDs.

The rise of the two classes of models comes from the diversity of
observed SNe Ia. In the literature, a number of explosion models have
been proposed to explain the normal and peculiar SNe Ia. For the
Ch-mass WD, representative models include the pure turbulent
deflagration model (PTD) \cite{Nomoto1976, Nomoto1984, Livne1993,
  Reinecke1999b, Reinecke2002a, Roepke2007, Ma2013, Leung2020SNIax},
PTD with deflagration-detonation transition \citep{Khokhlov1991a,
  Yamaoka1992, Iwamoto1999, Golombek2005, Roepke2007c, Fink2014, Leung2018Chand},
gravitationally confined detonation model \citep{Plewa2004,
  Meakin2009, Jordan2012, GarciaSenz2016, Seitenzahl2016} and
pulsation reverse detonation models \cite{Bravo2006a, Bravo2006b, 
  Bravo2009}. The subCh-mass WD models include the double-detonation
model \citep{Nomoto1982b, Fink2010, Shen2018, Tanikawa2018,
  Tanikawa2019, Leung2020subChand, Abigail2021, Shen2021, Gronow2021},
violent merger model \citep{Pakmor2011, Pakmor2012, Kromer2013,
  Tanikawa2015, Bear2018} and WD head-on collision model \citep{GarciaSenz2013,
  Kushnir2013, Papish2016}. On top of these, unconventional models
such as magnetized WDs\citep{Das2014}, super-Chandrasekhar mass WDs
\citep{Kamiya2012}, differentially rotating WDs\cite{Yoon2005, Hachisu2012} and
interaction with dark matter gravity\citep{Leung2015a,Chan2021} have
been proposed to explain some unusual SNe Ia.

\begin{table*}[]
    \tbl{Comparing essential features of Ch-mass and subCh-mass WDs}
    {\begin{tabular}{@{}llll@{}}
        \toprule
         &  unit & Ch-mass WD & subCh-mass WD \\ \hline
        mass & $M_{\odot}$ & 1.30 $- \geqslant$ 1.38 & 0.9 -- 1.2 \\
        central density & g cm$^{-3}$ & $10^9 - 10^{10}$ & $10^7$ -- $10^8$ \\ \hline
        composition & & $^{12}$C+$^{16}$O+$^{22}$Ne & core: $^{12}$C+$^{16}$O+$^{22}$Ne \\
        & & & envelope (env): $^{4}$He \\ \hline
        reaction & & subsonic deflagration & supersonic detonation \\
        first site & & (near-)center & off-center (He-env) \\ \hline
        
    \end{tabular}}
    \label{table:compare_WDs}
\end{table*}

The study of SNe Ia as explosions of (sub)Ch-mass WDs is often linked
to the open question about the progenitors of SNe Ia: the single
degenerate (SD) vs. the double degenerate (DD) scenario. The SD
scenario means that the primary WD develops its nuclear runaway by
mass accretion from its companion star, which can be a slightly
evolved main-sequence, a red-giant, or a He-star\citep{Hachisu2012, Nomoto2018}. The
DD scenario means that the primary WD triggers the runaway by
dynamical interaction with its companion WD.

\begin{figure*}
\includegraphics[width=2.4in]{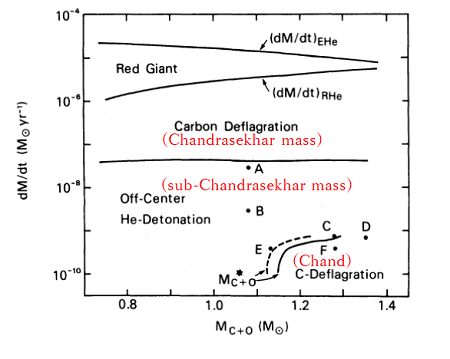}
\includegraphics[width=2.4in]{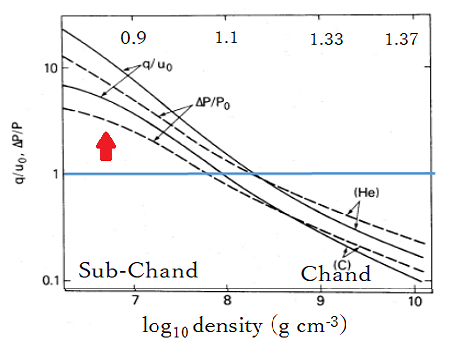}
\caption{(left panel) The final fate of the WD in the SD scenario with
  the mass accretion rate and the initial CO WD mass as parameters
  (derived and edited from Ref. \citenum{Nomoto1982a}).
 (right panel) The relative pressure change $\Delta P/P_0$ and
  relative internal energy change $q / u_0$ before and after nuclear
  runaway as a function of the matter density for the He-rich (solid
  line) and CO-rich (dashed line) matter
  (Ref. \citenum{Nomoto1982b}). The numbers on the top corresponds to
  the mass of the WD when the density corresponds to the central
  density of the WD. The red arrow indicates the relative pressure
  change of the CO-rich matter.}
\label{fig1:SD_outcome}
\end{figure*}

We remind that the question on whether SNe Ia develop from Ch-mass WDs
is not equivalent to arguing SNe Ia mainly develop in the SD
scenario. For example, in the SD scenario, when the WD explodes as an
SN Ia depends on the mass accretion rate from its companion star and
the WD initial mass (see the left panel of Figure
\ref{fig1:SD_outcome}). A WD having (1) a high mass accretion rate
above $\sim 10^{-9}~M_{\odot}$ yr$^{-1}$ or (2) having a low mass
accretion rate and a high initial mass $> 1.1~M_{\odot}$ is likely to
develop nuclear runaway in the Ch-mass limit. Otherwise, the WD is
more likely to explode as a subCh-mass WD \cite{Nomoto1982a}. Similar
features have been seen also for WDs in the DD scenario.

To understand why the C-deflagration is associated with the
Chandrasekhar mass WD, we show in the right panel of Figure
\ref{fig1:SD_outcome} the relative pressure change of the CO-rich
matter as a function of the matter density. During the thermonuclear
runaway, $^{12}$C and $^{16}$O burn to form iron-peak elements
peaked at $^{56}$Ni, releasing an amount of $\sim 10^{18}$ erg
g$^{-1}$. When the density is high ($\sim 10^{9}$ g cm$^{-3}$), the
electron degeneracy pressure dominates the matter pressure, and the
overall pressure becomes insensitive to its temperature. As a result,
the relative pressure jump decreases as the matter becomes
more degenerate. Without an abrupt pressure jump, the nuclear runaway in
the Ch-mass WD may not spontaneously trigger a shock wave and hence no
detonation may form. The hot matter may ignite $^{12}$C in the nearby cold
matter only by thermal conduction.

\begin{figure*}
\includegraphics[width=5in]{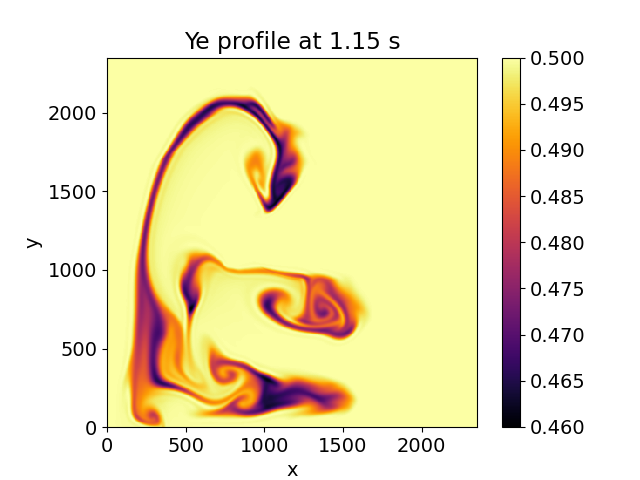}
\caption{ A snapshot of the electron fraction $Y_{\rm e}$ profile in a
  typical PTD model demonstrating simultaneously the Rayleigh-Taylor
  and Kelvin-Helmholtz instabilities due to interaction of turbulent
  fluid motion with the deflagration front. }
\label{fig:PTD_Ye}
\end{figure*}

Unlike the detonation, the subsonic deflagration is subject to
hydrodynamical instabilities such as the Rayleigh-Taylor (RT),
Kelvin-Helmholtz (KH) and Landau-Derrieus \cite{Niemeyer1995a,
  Bell2004a, Bell2004b} instabilities. The analytic model suggests
that the buoyancy force can drive the early flame away from the
center\cite{Fisher2015}.  In Figure \ref{fig:PTD_Ye} we plot the
electron fraction $Y_{\rm e}$ profile of a canonical PTD model where
the deflagration has quenched after the expansion of the WD. The
$Y_{\rm e}$ profile is a useful scalar for tracking how the fluid
elements move inside the star. We observe the elongated ``mushroom''
shape as features of the RT-instabilities and the spiral along and
inside the ``mushrooms'' as features of the KH-instabilities.

However, a WD may not naturally explode if there is only a slow subsonic
nuclear flame because the WD expands and quenches the flame before the
whole WD is burnt\citep{Nomoto1976, Reinecke2002a}. To alleviate this
issue, a deflagration-detonation transition\citep{Khokhlov1991a} and a
flame-acceleration scheme\citep{Nomoto1984, Niemeyer1995b,
  Woosley1997} have been proposed for assisting nuclear burning to
spread around the entire WD before the WD expands. 

\section{Typical Type Ia Supernova Explosion}

Both the Ch-mass and subCh-mass WD models have their individual
strengths and concerns, despite both of them can reproduce the
observed features of normal SNe Ia\citep{Graur2016, Shappee2017,
  Mori2018}, including the Philip's relation \citep{Kasen2009,
  Shen2021}. For example, the Ch-mass model can produce Mn with an
amount consistent with the solar abundance\citep{Seitenzahl2013b},
while the subCh-mass models do not produce a significant amount of Mn.
But the DDT mechanism remains a matter of debate whether or not the
turbulence is sufficient to pre-condition the CO rich
matter\citep{Lisewski2000, Roepke2007b, Woosley2009, Fisher2015, Fenn2016, 
  Brooker2021}.

\subsection{Typical Explosion Mechanism of Ch-mass and subCh-mass Models}

We now examine the typical explosion mechanism in both the Ch-mass and
subCh-mass WDs. Even though we have described a number of explosion
mechanisms in the previous section, in general they are only different
by the progenitor or the initial explosion kernel. The underlying
mechanism, namely the deflagration and detonation, remains
unchanged. Here we examine how the WD explodes accordingly.

\begin{figure*}
\includegraphics[width=2.4in]{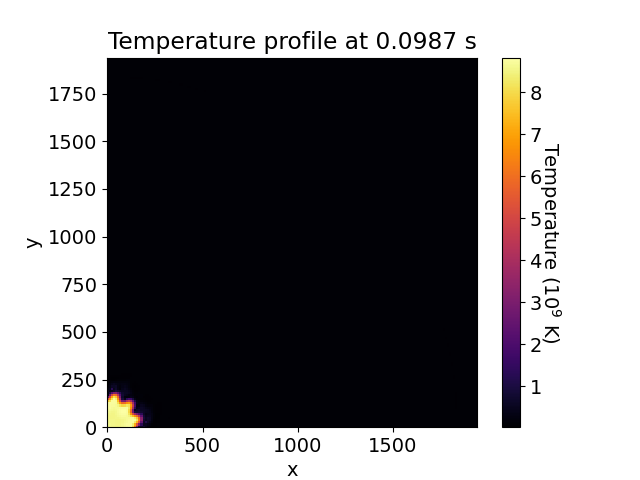}
\includegraphics[width=2.4in]{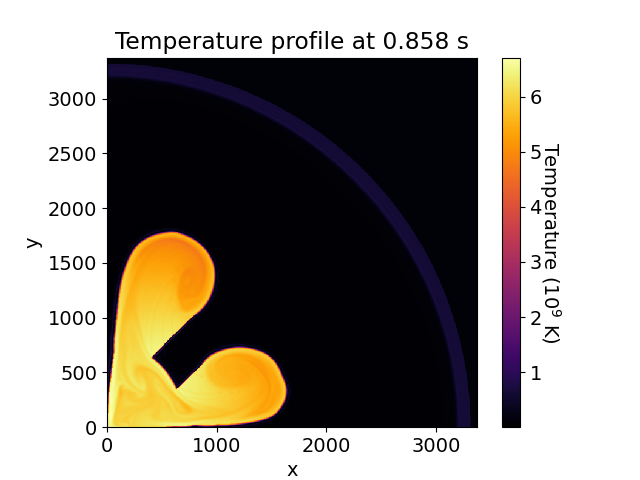}
\caption{(left panel) The initial temperature profile of the quadrant
  cross-section in a typical Ch-mass model using the PTD model with
  DDT for an initial mass $M = 1.37~M_{\odot}$, metallicity $Z =
  0.02$, and a ``three-finger'' initial flame
  kernel\citep{Leung2018Chand}. (top right panel) Same as the top left
  panel when the DDT is assumed to be triggered.}
\label{fig:Chand_temp_plot1}
\end{figure*}

\begin{figure*}
\includegraphics[width=2.4in]{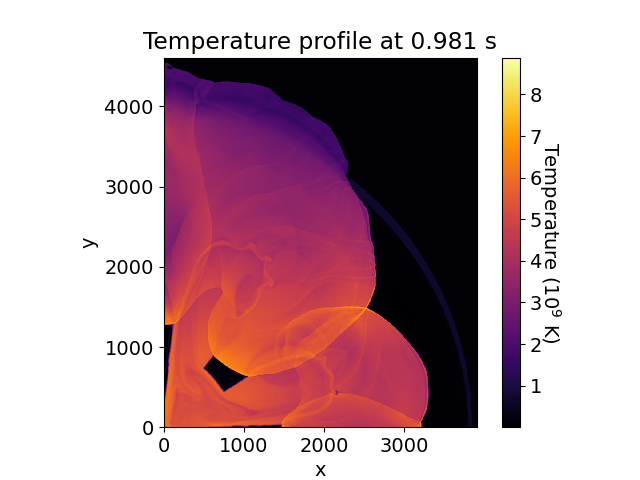}
\includegraphics[width=2.4in]{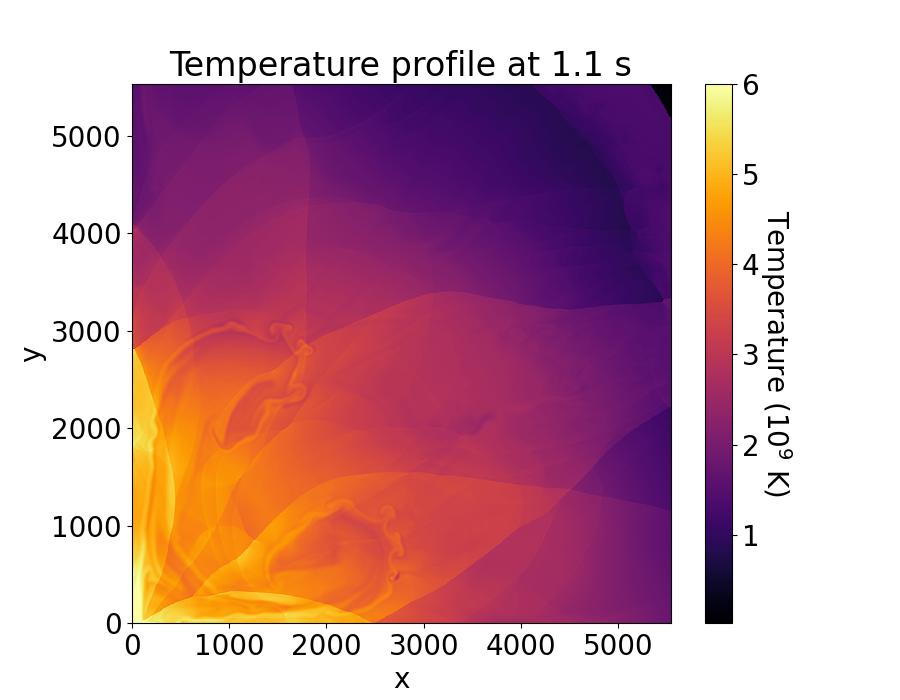}
\caption{(left panel) Same as Figure \ref{fig:Chand_temp_plot1} but
  during the detonation phase. (right panel) Near complete disruption
  of the WD.}
\label{fig:Chand_temp_plot2}
\end{figure*}

In Figures \ref{fig:Chand_temp_plot1} and \ref{fig:Chand_temp_plot2}
we plot the temperature profiles of the representative Ch-mass WD
explosion using the PTD model with DDT for a WD of 1.37 $M_{\odot}$,
metallicity $Z = $ 0.02 and a $c3$ deflagration
kernel\citep{Leung2018Chand} based on two-dimensional
simulations\citep{Leung2015b}. The WD is burnt by subsonic flame for
around 1 s, consuming about $\sim 30\%$ of the CO-rich matter in
mass. After that, DDT is assumed to take place and the remaining matter is burnt
within $\sim 0.1$ s.  Eventually, the WD undergoes homologous expansion which
quenches both deflagration and detonation.

\begin{figure*}
\includegraphics[width=2.3in]{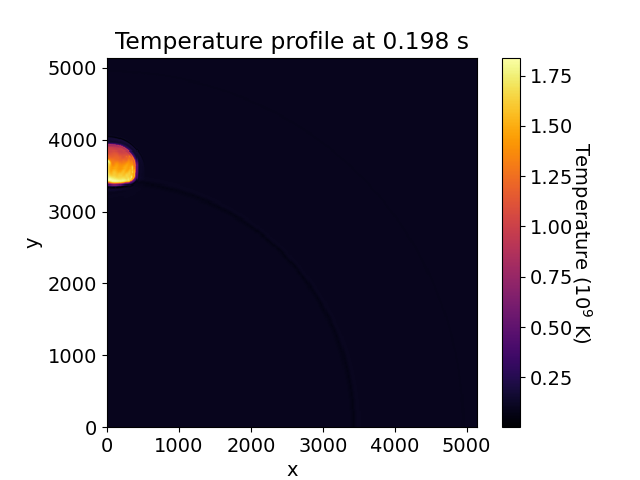}
\includegraphics[width=2.3in]{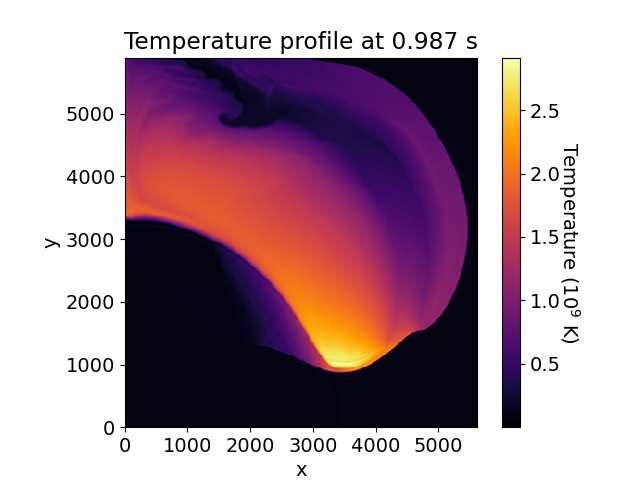}
\caption{(left panel) The initial temperature profile of a typical
  subCh-mass model using the double detonation model with the initial
  mass $M = 1.10~M_{\odot}$, $Z = 0.02$, and a ``single bubble''
  initial detonation kernel\citep{Leung2020subChand}. (right panel)
  Same as the top left panel but during the amplification of the He
  detonation.}
\label{fig:subCh_temp_plot1}
\end{figure*}

\begin{figure*}
\includegraphics[width=2.4in]{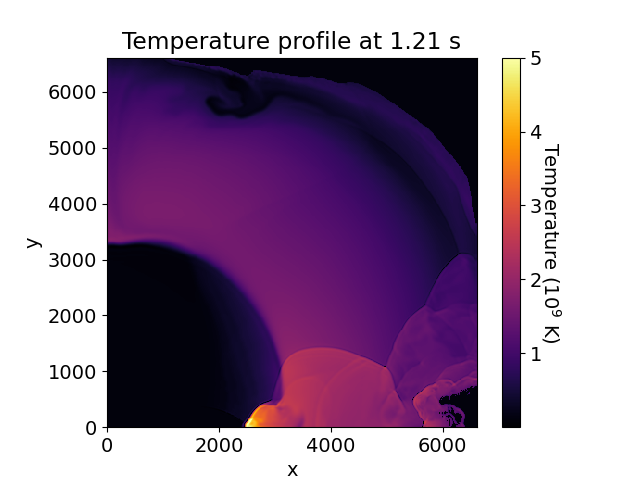}
\includegraphics[width=2.4in]{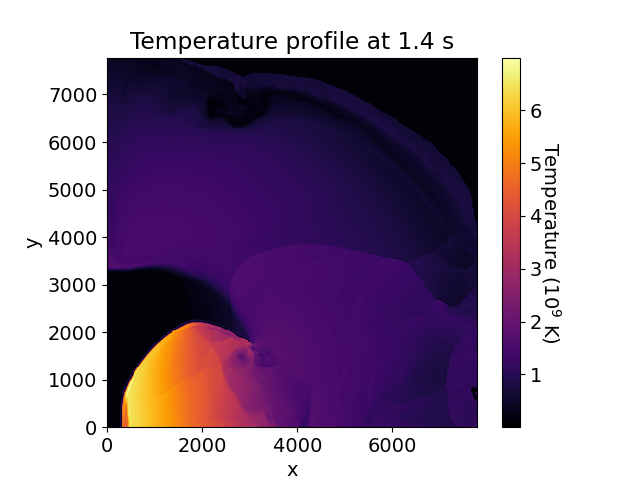}
\caption{(left panel) Same as Figure \ref{fig:subCh_temp_plot1} but
  during the onset of the C-detonation. (right panel) Same as the left
  panel but during the C-detonation phase.}
\label{fig:subCh_temp_plot2}
\end{figure*}

In Figures \ref{fig:subCh_temp_plot1} and \ref{fig:subCh_temp_plot2}
we plot similar profiles to Figures
\ref{fig:Chand_temp_plot1}-\ref{fig:Chand_temp_plot2} but for the
subCh-mass model with the initial mass $1.10~M_{\odot}$, $Z = 0.02$ and
a single He-detonation bubble\cite{Leung2020subChand}. In the first 1
s, the detonation burns the He-rich matter along the envelope. The
detonation strength increases during the collision, which creates a
shock that penetrates into the CO-core. This creates the C-detonation
which later disrupts the entire WD.

\subsection{General Thermodynamical Features}

\begin{table}[]
    \tbl{Major isotopes of iron-peak elements and their corresponding electron fraction}
    {\begin{tabular}{@{}lllllllllll@{}}
        \toprule
        \hline
        Isotope &  $^{54}$Fe & $^{55}$Mn & $^{55}$Fe & $^{55}$Fe & $^{56}$Fe & $^{56}$Co & $^{56}$Ni & 
        $^{57}$Fe & $^{58}$Ni & $^{60}$Ni\\ \hline
        Ye & 0.481 & 0.454 & 0.472 & 0.490 & 0.464 & 0.482 & 0.500 & 0.456 & 0.483 & 0.467 \\
        \hline
    \end{tabular}}
    \label{table:isotope_Ye}
\end{table}

Typical multi-dimensional SN Ia simulations solve the Eulerian
hydrodynamics equations with a simplified nuclear reaction network. To
obtain the detailed chemical features of the explosion, a passive
tracer particle scheme\citep{Shigehiro1998, Travaglio2004,
  Seitenzahl2010, Townsley2016} is necessary. This scheme allocates a
number of Lagrangian tracers to follow the fluid motion. The notation
"passive" means that the tracers do not affect the fluid motion; they
only record the thermodynamical condition along their trajectories.

The tracer particles record $(\rho(t), T(t))$ as a Lagrangian fluid
packet along its path for reconstructing the exact chemical
abundances. For SNe Ia, the trajectory is less convoluted that its peak
density and temperature $(\rho_{\rm peak}, T_{\rm peak})$ can
characterize the typical nucleosynthesis features inside the
tracer. We make numerical experiments to show how various
nucleosynthesis quantities depend on the parameters $(\rho_{\rm peak},
T_{\rm peak})$ parameter space.

We assume that the tracers start from given $(\rho_{\rm peak}, T_{\rm
  peak})$ and then adiabatically expand. The expansion timescale is
chosen according to the typical explosion energy $10^{51}$ erg. The
nuclear reactions are computed using the 495-isotope
network\citep{Timmes1999}.

\begin{figure*}
\includegraphics[width=2.4in]{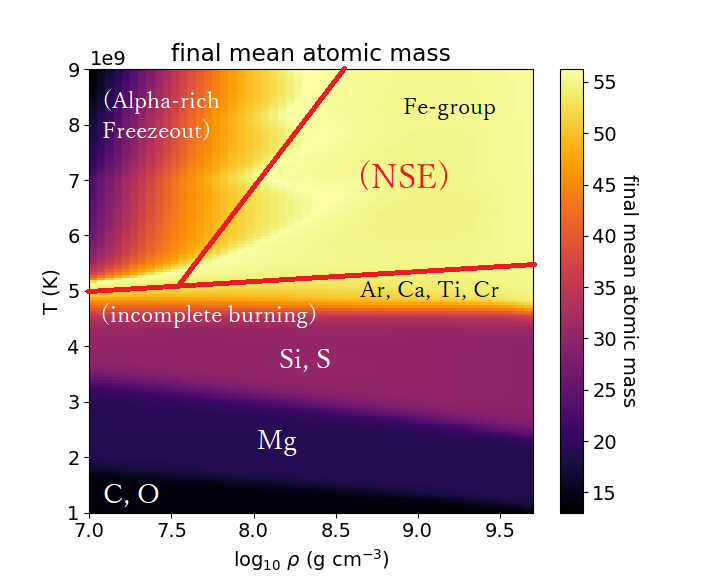}
\includegraphics[width=2.4in]{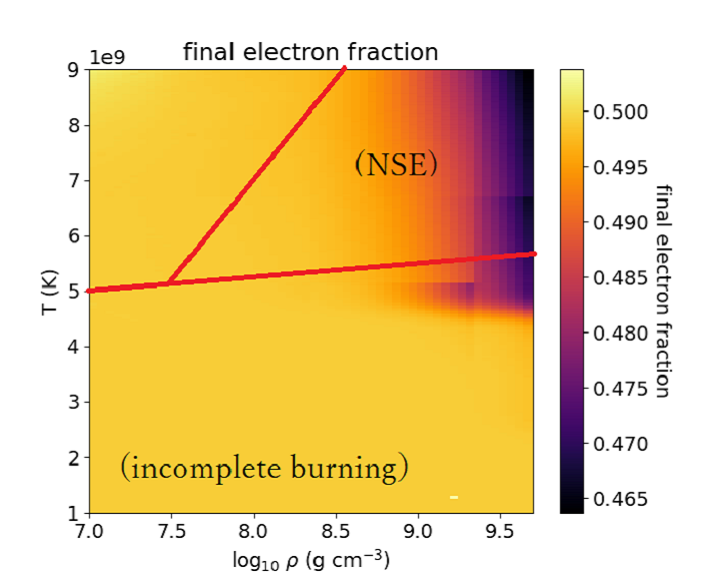}
\caption{(left panel) The final mean atomic mass number $\bar{A}$ of
  the tracer particles starting from different $\rho_{\rm peak}$ and
  $T_{\rm peak}$ (in units of 10$^9$ K). (right panel) Same as the left panel, but for the
  final electron fraction $Y_{\rm e}$ of the tracer.  }
\label{fig:thermo_traj1}
\end{figure*}

\begin{figure*}
\includegraphics[width=2.4in]{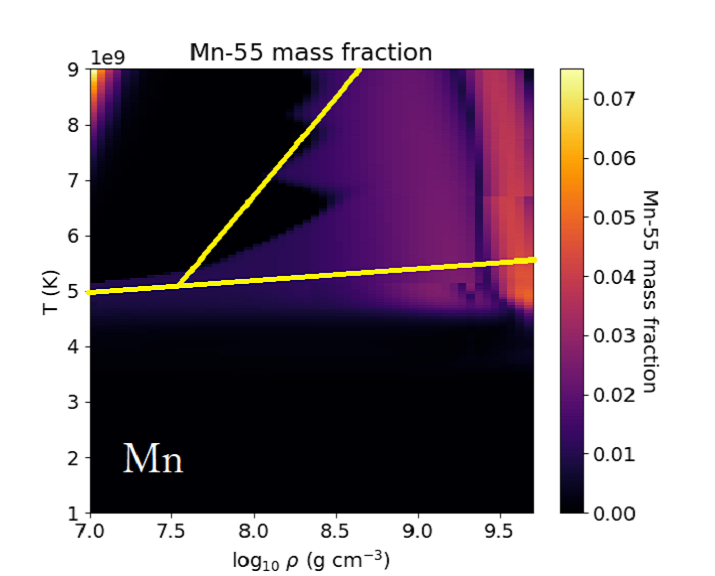}
\includegraphics[width=2.4in]{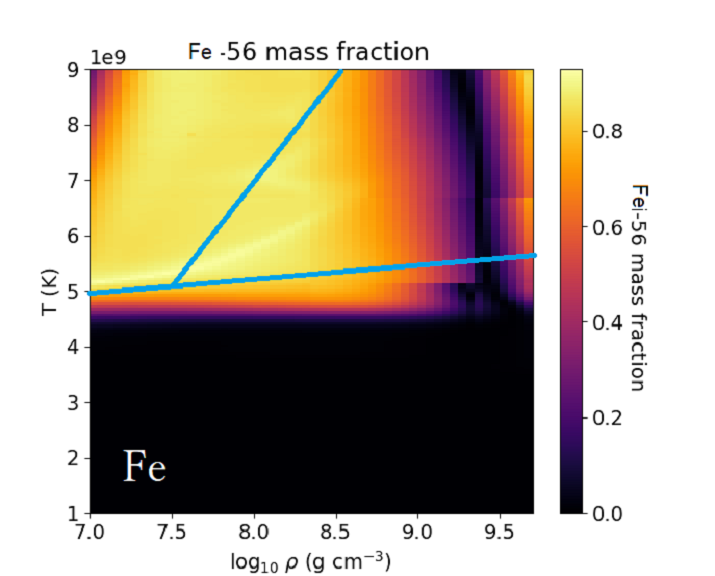}
\caption{(left panel) Same as Figure \ref{fig:thermo_traj1}, but for
  the final mass fraction of stable $^{55}$Mn. (right panel) Same as
  the left panel but for the final mass fraction of stable $^{56}$Fe.}
\label{fig:thermo_traj2}
\end{figure*}

In Figures \ref{fig:thermo_traj1} and \ref{fig:thermo_traj2} we plot
the final mean atomic number $\bar{A}$, $Y_{\rm e}$, asymptotic mass
fraction of $^{55}$Mn and $^{56}$Fe for tracers under different
initial conditions. The region is divided into three
regions\citep{Thielemann1986, Lach2020}. The low-$\rho_{\rm peak}$ region corresponds
to the incomplete Si-burning regime, where the nuclear reaction
terminates before reaching Fe-group elements, such as Si, S, Ar and so
on. The high-$T_{\rm peak}$ (in units of 10$^9$ K) and low-$\rho_{\rm peak}$ region
corresponds to the $\alpha$-rich freezeout regime. As the name
suggests, the nuclear reaction is confined to be along the
$\alpha$-chain from $^{12}$C to $^{56}$Ni. The high-$T_{\rm peak}$ and
high-$\rho_{\rm peak}$ region corresponds to the nuclear statistical
equilibrium (NSE) regime. This regime plays an important role in the
Ch-mass WD as it allows isotopes away from the $\alpha$-chain to form
through weak interaction (electron capture).

As the $Y_{\rm e}$-profile indicates, the NSE zone is also the region
where matter with $Y_{\rm e} < 0.5$ can be formed. The low $Y_{\rm e}$
environment is vital for forming the parents of $^{55}$Mn (see Table
\ref{table:isotope_Ye} for the representative $Y_{\rm e}$ for the
major neutron-rich isotopes of iron-peak elements). The $^{55}$Mn
profile also shows that the NSE zone is the primary site for
generating a significant amount of stable $^{55}$Mn after decay. On
the other hand, $^{56}$Fe is mostly formed in the $\alpha$-rich
freezeout and NSE ($Y_{\rm e} \approx 0.5$) regions.

\subsection{Thermodynamical Trajectories of SN Ia Models}

\begin{figure*}
\includegraphics[width=2.4in]{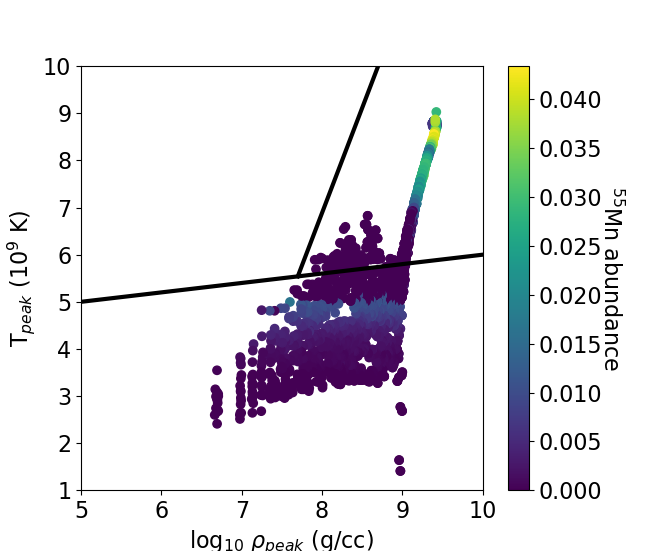}
\includegraphics[width=2.4in]{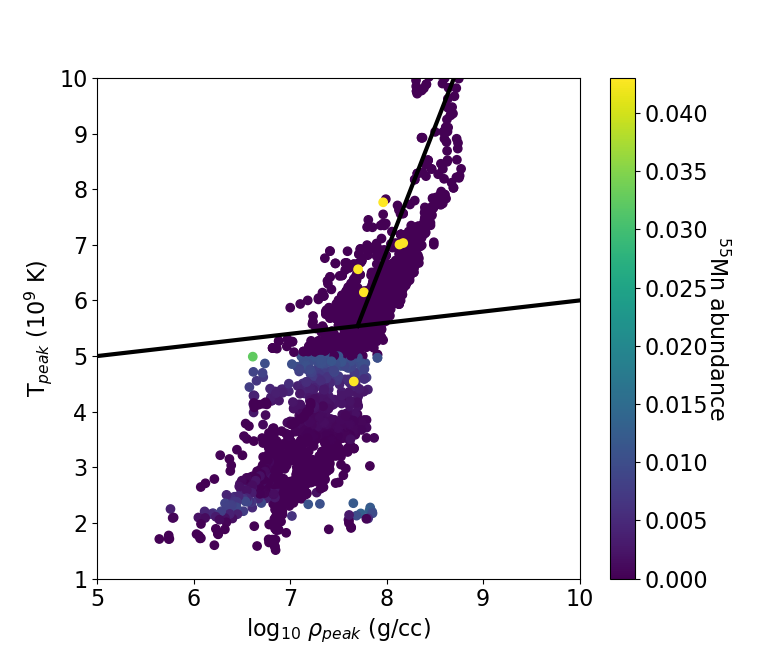}
\caption{(left panel) The thermodynamical trajectories of tracer
  particles of the Ch-mass model with the colour being the asymptotic
  $^{55}$Mn abundance. Same as the left panel but for the subCh-mass
  model.}
\label{fig:thermo_traj_SNIa}
\end{figure*}

Having explored which thermodynamical parameter space is responsible
for iron-peak elements, we show in Figure \ref{fig:thermo_traj_SNIa}
the thermodynamics trajectories of tracers obtained from the typical
Ch-mass and subCh-mass models. The chemical abundance of each tracer
is directly computed according to its individual $(\rho,T)$ time
evolution.

The Ch-mass model (left panel) has two distinctive parts: the high
density thin tail and the thick body at low density. At high density
($\rho_{\rm peak} > 10^9$ g cm$^{-3}$), the tracers are in the NSE
regime and have a significantly higher $^{55}$Mn and low fluctuations
in $T_{\rm peak}$ for the same $\rho_{\rm peak}$. These are the
tracers burnt by the subsonic deflagration. The absence of shock
ensures that nuclear burning does not generate strong acoustic
waves. On the other hand, the majority of tracers burnt by the
detonation undergo incomplete Si-burning. The aspherical explosion
allows tracers with the same initial mass coordinate to be burnt at a
range of time. This leads to a wide temperature range for the same
$\rho_{\rm peak}$. There is also a narrow band of tracers for $7 <
\log_{\rm 10} \rho_{\rm peak} < 9$ and $T_{\rm peak} \approx 5 \times
10^9$ K also responsible for synthesizing a small fraction of
$^{55}$Mn.

The subCh-mass model (right panel) has a uniform structure where the
$T_{\rm peak}$ scales with $\rho_{\rm peak}$ with some
fluctuations. Only a small part of tracers reaches the NSE regime but
their density is not high enough for the $^{55}$Mn synthesis. There is
also a narrow band of tracers containing $^{55}$Mn by the synthesis of
$^{55}$Co. In general the global $^{55}$Mn in the subCh-mass model is
lower than that of the Ch-mass model.

\subsection{Typical Nucleosynthesis in Ch-mass and subCh-mass Models}

\begin{figure*}
\includegraphics[width=4.5in]{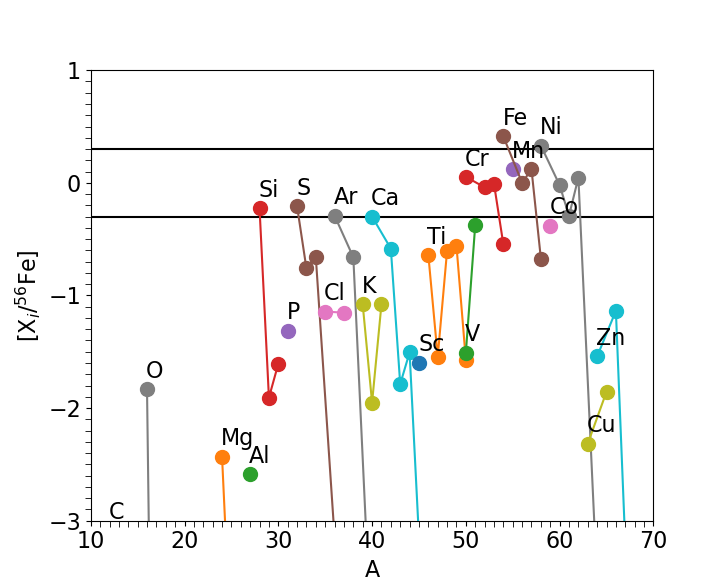}
\includegraphics[width=4.5in]{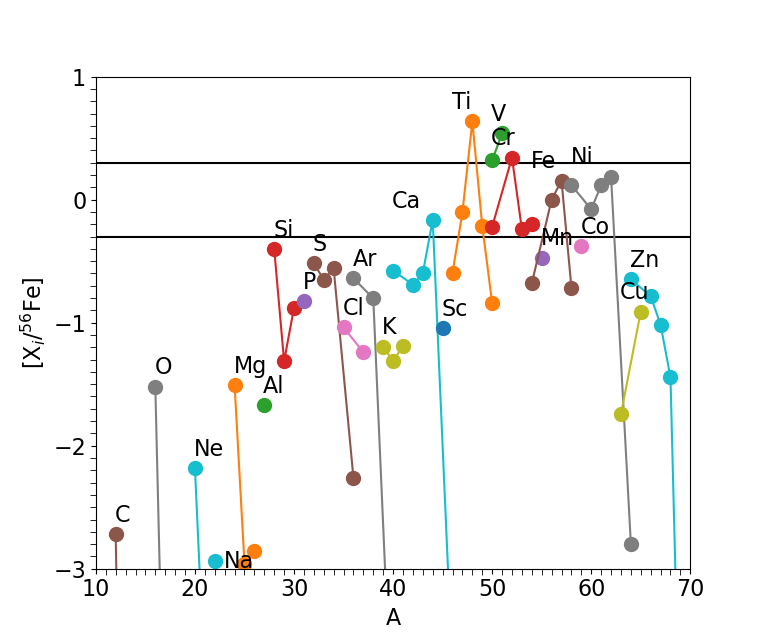}
\caption{(top panel) The final chemical abundance pattern of the
  typical Ch-mass WD\cite{Leung2018Chand} assuming the aspherical
  explosion. (bottom panel) Same as the top panel but for the typical
  subCh-mass WD\cite{Leung2020subChand}. $[X_i/^{56}$Fe] $= \log_{10}
  [(X_i/^{56}$Fe) / ($X_i/^{56}$Fe)$_{\odot}$]. The two horizontal
  lines correspond to 50 \% and 200 \% of the solar value.}
\label{fig:abundances}
\end{figure*}

Now we have examined the thermodynamical differences between the
Ch-mass and subCh-mass WDs. In Figure \ref{fig:abundances} we compare
the qualitative differences in the nucleosynthesis pattern.

Both Ch-mass and subCh-mass WDs share some common features. They are
responsible for the production of intermediate mass elements (IMEs)
from Si to Ca, and the iron-peak elements from Ti to Ni. Odd number
elements of IMEs are underproduced in SNe Ia. Some individual features
allow us to distinguish the two models. (1) The aspherical explosion
of the subCh-mass model can lead to signatures of strong Ti, V and
Cr. (2) Mn is well-produced in the Ch-mass model but not in the
subCh-mass model.

\section{Applications of Nucleosynthesis}

We have surveyed the major differences of the nucleosynthetic
signature between the Ch-mass and subCh-mass WDs. Comparisons with
observational data allow us to understand the progenitors of observed
SNe Ia, which directly constrains the modeling. We can compare the
optical signatures directly (i.e., light curves and spectra) by matching
the radiative transfer model with SN Ia data\citep{Kromer2013,
  Graur2016}. One can also extract the chemical abundances from the
spectra, and compare with nucleosynthetic results\citep{Mori2018,
  Leung2021SN2014J}. We shall focus on the latter method here.

\subsection{Supernova Remnant Sagittarius A East}

\begin{figure*}
\includegraphics[width=2.4in]{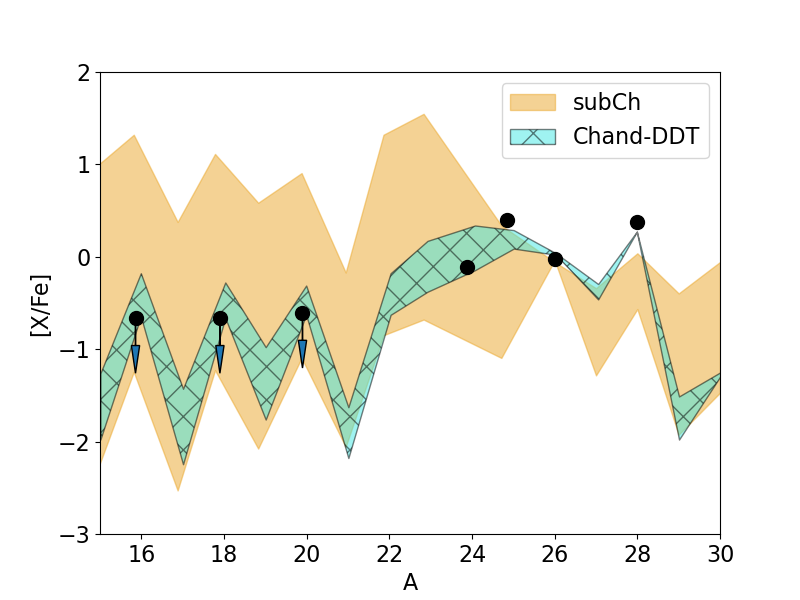}
\includegraphics[width=2.4in]{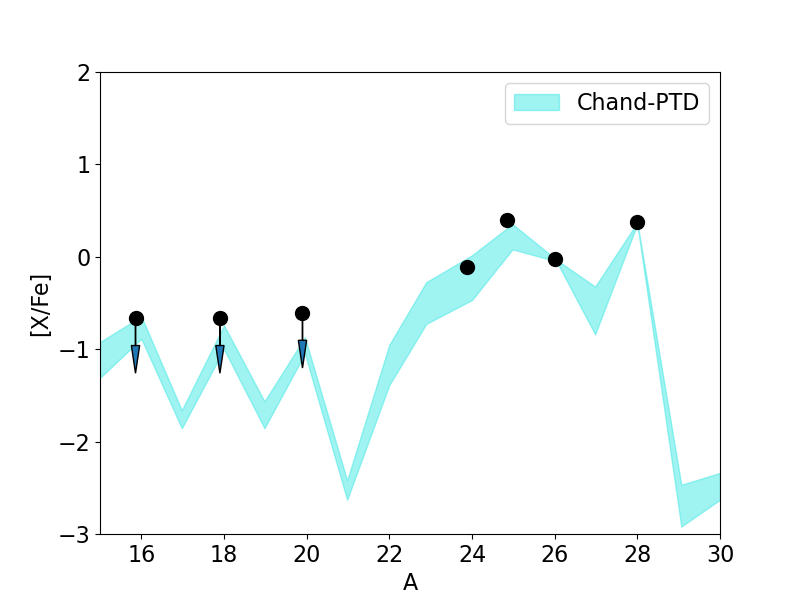}
\caption{(left panel) The chemical abundance pattern of the supernova remnant (SNR)
  Sagittarius A (from Ref. \citenum{Zhou2021}) for the data points
  compared with those of the subCh-mass\citep{Leung2020subChand} and
  Ch-mass DDT\citep{Leung2018Chand} models shown by 
  the shaded regions.  (right panel) Same as the
  left panel but for the Ch-mass PTD\citep{Leung2020SNIax} models.}
\label{fig:SagA}
\end{figure*}

Within thousand years after the SN explosion, the shock-heated gas
remains observable in the X-ray band, where the spectra reveal the
metal composition inside the ejecta. Such a technique has been applied
to the study of galactic supernova remnants (SNRs) including Tycho\citep{Park2013},
Kepler\citep{Yamaguchi2017} and N103B\citep{Yamaguchi2021}.

In Ref. \citenum{Zhou2021} the SNR in Sagittarius A (Sgr A) East
(G0.0+0.0) is observed based on the X-ray data taken by the
\textit{Chandra} telescope.
The observed abundance ratios relative to Fe (with respect to the solar ratios)
[Xi/Fe] are shown in Figure \ref{fig:SagA}. The SNR features sub-solar
intermediate mass elements (IMEs) and slightly super-solar iron-peak elements (Cr, Mn, and
Ni).

The sub-solar IMEs exclude the possibility of associating a
core-collapse SN as the origin of this remnant. On the left panel, the
abundances of two distinctive classes of models, the subCh-mass and
Ch-mass DDT models are plotted. The model uncertainties are shown by the
shaded area. The subCh-mass models clearly overproduce the IME.
Among the Ch-mass DDT models, the model that produces enough Mn and Ni
overproduces Cr and the IME.  There is a model whose Cr and Ni 
are consistent with the data points and IMEs are marginal, but its
Mn is too small.

The Ch-mass PTD model
(i.e., no DDT) with the initial central density of $\sim 5 \times 10^9$ g cm$^{-3}$
is  shown to be compatible with the data (right
panel of Figure \ref{fig:SagA}).
[Such a high central density is realized in the rotating WD model
  \citep{Benvenuto2015}.]
Note that this Ch-mass PTD model can well-explain the observed features of
SNe Iax.  Thus this object is the first identified
SN Iax in the Milky Way Galaxy observed as SNR. This example also
shows how the abundance guides us to identify the explosion mechanism.

\subsection{Supernova Remnant 3C 397}

\begin{figure*}
\includegraphics[width=2.4in]{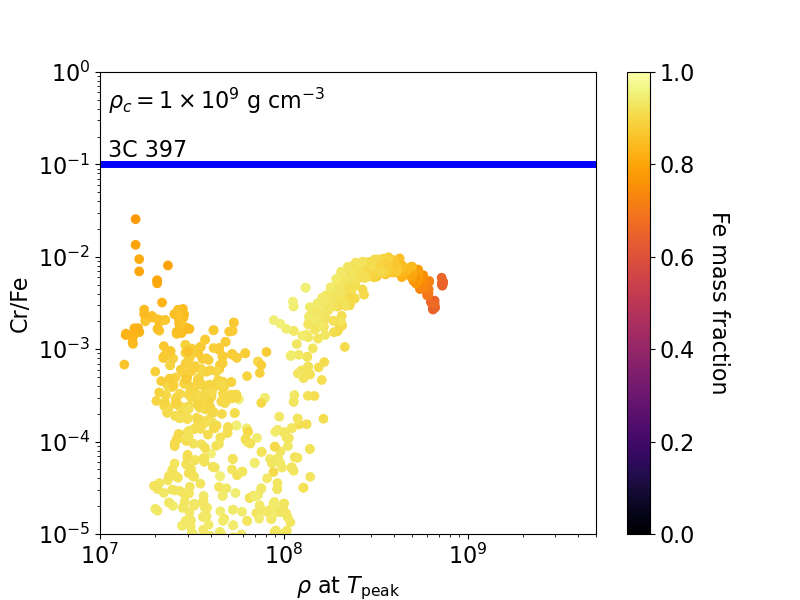}
\includegraphics[width=2.4in]{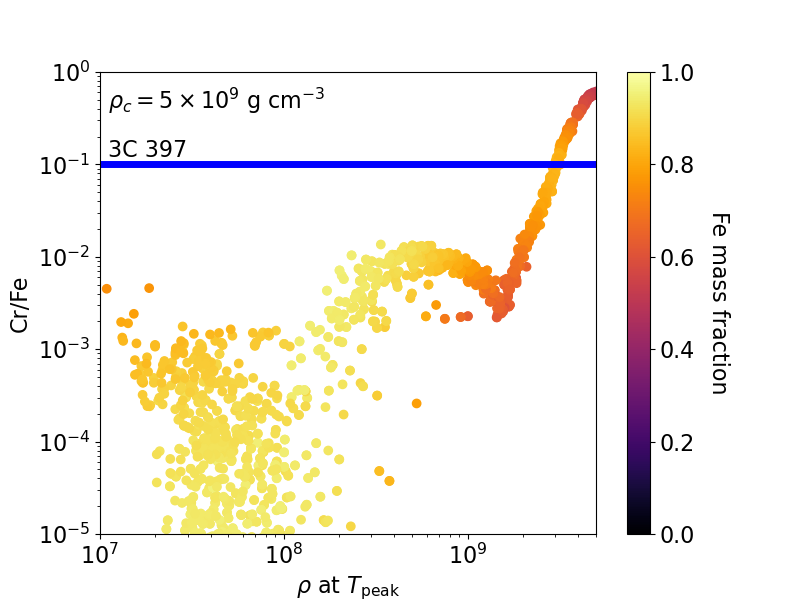}
\caption{(left panel) The Cr/Fe distribution of the tracers taken from
  the Ch-mass model with the initial central density $1 \times 10^9$ g
  cm$^{-3}$\citep{Leung2018Chand}. The color represents the tracer Fe
  mass fraction. The horizontal line is the measured value in SNR 3C
  397 from Ref. \citenum{Ohshiro2021}. (right panel) Same as the left
  panel but for the model with the initial central density $5 \times 10^9$ g cm$^{-3}$}
\label{fig:3C397}
\end{figure*}

The SNR 3C 397 is a nearby object (8 kpc) on the
galactic plane. Its close distance allows astronomers to extract the
spectra from individual parts similar to Sgr A. This object features a
high Mn/Ni ratio, which is a key evidence of the Ch-mass explosion
\citep{Yamaguchi2015}.

In a recent observation using the \textit{XMM-Newton} telescope, the
spectra from the South and West hot blobs are measured, which give the
constraint on the Cr/Fe mass ratio $\sim 0.106
\pm^{0.011}_{0.009}$\citep{Ohshiro2021}. The high value is used to
distinguish the explosion progenitor shown in Figure
\ref{fig:3C397}. By comparing the tracer in different Ch-mass models,
it becomes clear that the low-mass model ($\rho_{c} = 1 \times 10^9$ g
cm$^{-3}$) does not have tracers reaching the observed high
value. Meanwhile the high density tail in the high-mass model
($\rho_{c} = 5 \times 10^9$ g cm$^{-3}$) has tracers crossing the
expected value. This provides a strong indication that this object is
the explosion of the high-mass Ch-mass WD. This also demonstrates how a
precise measurement of element abundance ratios can guide us to select
the potential progenitor.

\subsection{Milky Way Galaxy}

In the last two sections we have shown how the SNR abundance
determines its progenitor and the explosion mechanism. While there is
no distinctive SNR showing chemical abundances exclusively for
subCh-mass WD models, it is possible that a large sample size is
needed to understand the distribution of each model. To understand the
SN Ia explosion globally, we need the chemical abundances from a
larger system, for example, the Milky Way Galaxy. The elements ejected
by supernovae become the building block of the next-generation
stars\citep{Matteucci2001}. The surface abundance of stars in the
solar neighbourhood may thus indicate how much each element is ejected
by generations of SNe Ia.

\begin{figure*}
\includegraphics[width=5in]{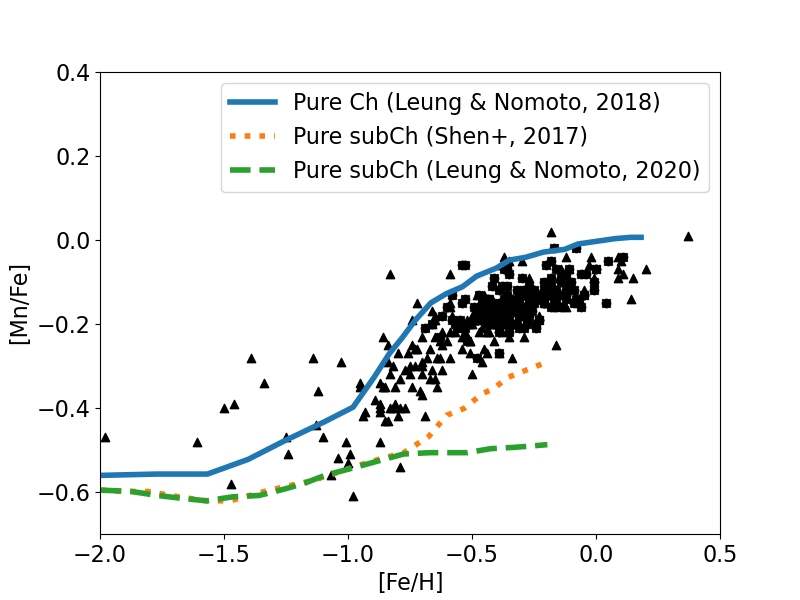}
\caption{The [Mn/Fe] against metallicity [Fe/H] for the galactic
  chemical evolution models taken from
  Ref. \citenum{Kobayashi2020}. Solid lines come from theoretical
  models assuming pure Ch-mass and subCh-mass explosion history. Data
  points are the stellar abundances from the solar neighbourhood
  \cite{Reddy2003, Reddy2006, Feltzing2007}.}
\label{fig:GCE}
\end{figure*}

In Ref. \citenum{Kobayashi2020} the galactic chemical evolution model
is computed with supernova abundance patterns taken from
literature. The Mn/Fe evolution is plotted in Figure
\ref{fig:GCE}. Two contrasting classes of models are shown, one
assuming the pure Ch-mass WD explosion, and the other two assuming
pure subCh-mass WD. To reproduce the trend as well as the magnitude of
the data, a non-negligible fraction of the Ch-mass WD is necessary.

We remark that the supernova history can be strongly dependent on the
galaxy evolution history. Some galaxies (e.g., Sculptor dwarf
spheroidal galaxy) have a low Mn/Fe ratio that indicates the dominance
of the subCh-mass WD explosion in their evolution histories
\citep{Kirby2019, delosReyes}. Meanwhile, some early rise of [Mn/Fe]
in this subclass of galaxies can be a result of the Ch-mass SN Iax
explosion \citep{Kobayashi2015}.

\subsection{Perseus Galactic Cluster}

The Milky Way Galaxy can provide a detailed reference in how
generations of stars contributes to the cosmic metal
enrichment. However, large N-body simulations suggest that each galaxy
is unique in their evolution history. To understand how each supernova
model contributes in the cosmic scale, data from an even larger system
is important to average out the statistical fluctuations of individual
galaxies.

In Ref. \citenum{Simionescu2019} the X-ray spectra of the Perseus
Cluster is studied by the \textit{Hitomi} telescope. The highly
resolved spectral lines provide the abundance measurement with 
uncertainties down to $\sim 10\%$. The high precision can distinguish
supernova models and mechanisms explicitly. The fitting using SN Ia
and CCSN models from literature is shown in Table
\ref{table:Perseus}. The best-fit model is found to be the scenario
assuming pure Ch-mass WD explosion. If the fraction of the Ch-mass WD is
relaxed as a model parameter, the expected Ch-mass WD still
contributes about 10 -- 40 \% of the SN Ia population, depending on
the exact CCSN models.

\begin{table}[]
  \tbl{Models assuming different stellar and supernova models and
    their corresponding (Ch-mass) SN Ia rates (data taken from
    [Ref. \citenum{Simionescu2019}])}
 {\begin{tabular}{@{}llll@{}}
        \toprule
        \hline
        Model & $f_{\rm Ia}$  & $f_{\rm Chand}$ & $\chi^2$ \\ 
        pure Ch-mass[Ref. \citenum{Leung2018Chand}]+ CCSN[Ref. \citenum{Nomoto2013}] & 0.21 $\pm$ 0.02 & 
        N/A & 11.78 \\ \hline
        Ch-mass[Ref. \citenum{Seitenzahl2013a}] + subCh-mass[Ref. \citenum{Shen2018}] + CCSN[\citenum{Nomoto2013}] & 0.25 $\pm$ 0.06 & 0.36 $\pm$ 0.14 & 23.96 \\
        Ch-mass[Ref. \citenum{Seitenzahl2013a}] + subCh-mass[Ref. \citenum{Shen2018}] + CCSN[\citenum{Sukhbold2016}] & 0.38 $\pm$ 0.06 & 0.09 $\pm$ 0.09 & 15.73 \\ \hline
        
        \hline
    \end{tabular}}
    \label{table:Perseus}
\end{table}

\section{Conclusion}

In this review article we have presented the physical background about
the Ch-mass and subCh-mass WD models as the SN Ia explosion progenitors. We
discussed the differences in their explosion mechanisms and their
associated nucleosynthetic signatures. We have also demonstrated how
the chemical abundances of SNRs, Milky Way Galaxy, and galactic
clusters can help us distinguish (1) the individual SN explosion
scenario and (2) the relative importance of each explosion model.

Nucleosynthesis will remain an important subject in the future
supernova study thanks to observational projects such as XRISM (X-Ray
Imaging and Spectroscopy Mission). Given the power of resolving
spectral lines as its predecessor \textit{Hitomi}, we can anticipate
that the high quality spectral data, and hence the precise chemical
abundance measurements, will shed light on supernova models to an
unprecedented accuracy.

\section*{Acknowledgments}

S.C.L. thanks the session chairpersons Pilar-Ruiz Lapuente and Robert
Fisher for the invition to the introductory talk in the Marcel
Grossmann 16 Meeting.  S.C.L acknowledges support by NASA grants
HST-AR-15021.001-A and 80NSSC18K1017. K.N. has been supported by the
World Premier International Research Center Initiative (WPI
Initiative), MEXT, Japan, and JSPS KAKENHI Grant Numbers JP17K05382, 
JP20K04024, and JP21H04499.

\bibliographystyle{ws-procs961x669}
\bibliography{main}

\end{document}